\documentclass[11pt,a4paper]{article}
\usepackage{jcappub}

\def\be{\begin{equation}}
\def\ee{\end{equation}}
\def\beq{\begin{equation}}
\def\eeq{\end{equation}}
\def\bea{\begin{eqnarray}}
\def\eea{\end{eqnarray}}
\def\bml{\begin{subequations}}
\def\blea{\bml\begin{eqnarray}}
\def\elea{\end{eqnarray}\end{subequations}}

\usepackage{amsmath}

\newcommand{\Hub}{\mathcal{H}}
\newcommand{\B}{\mathbf}
\newcommand{\F}{\mathcal} 
\newcommand{\der}[2]{\frac{\partial#1}{\partial#2}}
\begin{document}
\title{Transport Equation for Nambu-Goto Strings}
\author{Daniel Schubring and}

\emailAdd{schub071@d.umn.edu}

\author{Vitaly Vanchurin}

\emailAdd{vvanchur@d.umn.edu}

\date{\today}

\affiliation{Department of Physics, University of Minnesota, Duluth, Minnesota, 55812}

\abstract{

We consider a covariant approach to coarse-graining a network of interacting Nambu-Goto strings. A transport equation is constructed for a spatially flat Friedmann universe. In Minkowski space and with no spatial dependence this model agrees with a previous model. Thus it likewise converges to an equilibrium with a factorizability property. We present an argument that this property does not depend on a `string chaos' assumption on the correlations between strings. And in contrast to the earlier model, this transport equation agrees with conservation equations for a fluid of strings derived from a different perspective.}
\maketitle

\tableofcontents
\section{Introduction}

Consider a very large network of one-dimensional objects (which we shall call strings) whose individual dynamics and interactions are described by some physical laws. These could be cosmic strings formed after a cosmological phase transition \cite{Cosmic}, fundamental strings near the Hagedorn temperature \cite{Fundamental}, topological strings in a nematic liquid crystal \cite{Crystal} or even more complicated objects such as polymer molecules \cite{Polymer}. If the number of bits of information required to specify a state of the system (or the number of degrees of freedom) is small, then one can try to simulate it on a computer (as in Refs. \cite{Numerical}), but what if the number of bits is very large? Can we say anything meaningful about such systems? 

The standard approach would be to make use of equilibrium statistical mechanics. However, this may fail for one or more of the following reasons: the system might not be in equilibrium, the ergodic hypothesis might not be obeyed, the dynamics might not be Hamiltonian, the partition function might diverge, etc. But if it is not equilibrium statistical mechanics, then what else can we do to describe the collective behavior of strings? There are at least two more options that are not completely unrelated and both involve coarse-graining of the strings on some scales: one based on the ideas of fluid mechanics \cite{Landau} and the other based on the methods of kinetic theory \cite{Huang}. 

The first option is to coarse-grain the network of strings and to treat it as a fluid described by a number of fields such as the energy density, velocity and tangent vector fields. Then by considering flows of conserved quantities (e.g. energy, momentum, tangent vectors, etc.) one can derive equations analogous to Euler equations for a fluid of particles that the coarse-grained fields must obey. The second option is to derive a transport equation, analogous to the Boltzmann transport equation,  for a distribution function of strings by considering the dynamics as well as interactions of individual strings. Then one can try to solve the transport equation, either analytically or numerically, to study the evolution of the systems towards an equilibrium or a steady state which may or may not be unique. 

Note that the two approaches are not completely independent and the equilibrium derived in the context of the kinetic theory can be used to simplify the fluid equations. For example, if the coarse-graining scales in the fluid description are  sufficiently large then one can assume that the smaller local subsystems are quickly driven towards a state of a local equilibrium. Of course the local equilibrium assumption is only valid on very large scales and is guaranteed to break down on small scales where the higher order correction to the fluid equations become important. Such corrections can not be obtained by considering conserved quantities in the fluid model, but can be obtained, for example, from a perturbative expansion of the transport equation around equilibrium.  

One should keep in mind that although the kinetic theory approach seems to be a lot more precise it often depends on additional assumptions (e.g. molecular chaos, string chaos, Markov property). Moreover, the transport equations are usually integro-differential equations that are very hard to solve. In contrast the fluid equations are only differential equations that are relatively easy to solve if the solutions exist. In reality the first order fluid equations (e.g. Euler equations) often develop shock waves, and even the second order fluid equations  (e.g. Navier-Stokes equations) may or may not have smooth solutions. The physically relevant quantities are rarely  discontinues and the break-down of solutions indicates that the fluid description is incomplete unless higher order corrections are included to smooth out the discontinuities. 

Both approaches were already implemented to study the dynamics of a network of interacting Nambu-Goto strings. The fluid analysis revealed that in addition to conserved currents corresponding to the symmetric energy-momentum tensor, there are conserved currents associated with an anti-symmmetric tensor, both of which lead to the first order fluid equations \cite{Fluid}. The kinetic theory of strings was developed by considering a distribution of velocities and tangent vectors of uncorrelated string segments \cite{OldKinetic, Kinetic}. In a homogenous limit the transport equation was derived and an $H$-theorem for strings was proved to show that the equilibrium distribution has a  factorizability property (to be discussed in details below) \cite{Kinetic}. Then, under assumption of a local equilibrium, the first order fluid equations take a particularly simple form \cite{Fluid}, but their solutions are likely develop shock waves. 

The next step could have been to expand the transport equation around equilibrium to obtain the high order corrections to the fluid equations, but the transport equation from Ref. \cite{Kinetic} does not seem to agree with the fluid equations from Ref. \cite{Fluid} even at the linear order. In this paper we will remove the discrepancy by providing a microscopic derivation of an inhomogeneous transport equation that is in a full agreement with the fluid equations derived in Ref. \cite{Fluid} as well as with the homogeneous transport equation derived in Ref. \cite{Kinetic}. 

This paper is organized as follows. In Sec. \ref{Sec:Timeslice} the time-slice coordinates are constructed and the corresponding equations of motion are derived. The kinetic theory is developed in Sec. \ref{Sec:Kinetic} and the equilibrium solutions of the transport equation are analyzed in Sec. \ref{Sec:Equilibrium}. The transport equation in the inhomogeneous limit and in the Friedmann universe is derived in Secs. \ref{Sec:Inhomogeneous} and \ref{Sec:Friedmann} respectively. The main results of the paper are summarized in Sec. \ref{Sec:Summary}.

\section{Worldsheet Coordinates}\label{Sec:Timeslice}
In the tangent space at any point on a worldsheet there are two distinct lightlike rays (which may coincide in the degenerate case). If there is some physical distinction which allows us to consistently define a right and left direction along the length of a string, we can call the left pointing ray $ A $, and the right pointing ray $ B $.

From these rays, we can choose vector fields $ \F{A}^\mu $ and $ \F{B}^\mu $ over the worldsheet. One natural choice is to choose the fields to be coordinate vectors, leading to light-cone coordinates $ \zeta^a $. The equations of motion of a Nambu-Goto string (with units chosen to set the string tension to one), are particularly simple in these coordinates:
\begin{align}
\F{B}^\lambda \nabla_\lambda \F{A}^\mu  = \F{A}^\lambda \nabla_\lambda \F{B}^\mu = 0.\label{eqconformal}
\end{align}
So $ \F{A} $ is parallel-transported along the string in the direction $ B $, and $ \F{B} $ is parallel-transported along the direction of $ A $.

There is still some gauge freedom left in choosing light-cone coordinates. In dealing with a network of many strings on which coordinates can be assigned independently on each, we must be careful to use gauge invariant quantities. One such example is the energy-momentum tensor, which can be written as a volume form over the worldsheet in a manifestly covariant way (see for instance \cite{Review}),
\begin{equation}
\tilde{T}^{\mu\nu} = h^{ab}\der{x^\mu}{\zeta^a}\der{x^\nu}{\zeta^b} \sqrt{-h}\,d^2\zeta.
\label{T}
\end{equation}
In light-cone coordinates this can be written,
\begin{align}
\tilde{T}^{\mu\nu} = 2\F{A}^{(\mu}\F{B}^{\nu)}\,d^2\zeta
\label{Tconformal}
\end{align}

\subsection{Time-Slice Coordinates}

In flat spacetime we can get rid of the remaining gauge freedom in light-cone coordinates by fixing the time components $ \F{A}^0=1 $ and $ \F{B}^0=1 $ everywhere. Even in general spacetime we can still choose unique vector fields $ A^\mu $ and $ B^\mu $ corresponding to the rays $ A $ and $ B $ which have a timelike component of one everywhere. From these, we can define timelike $ v^\mu $ and spacelike $ u^\mu $ as
\begin{align}
v^\mu \equiv \frac{1}{2}(B^\mu + A^\mu)\nonumber\\
u^\mu \equiv \frac{1}{2}(B^\mu - A^\mu)
\label{vu}
\end{align}
And since $ A^\mu $ and $ B^\mu $ are lightlike we have the relations:
\begin{align}
v^\lambda u_\lambda = 0\label{gauge1}\\
v^\lambda v_\lambda + u^\lambda u_\lambda = 0\label{gauge2}
\end{align}

However, $ v^\mu $ and $ u^\mu $ need not be the vectors of any coordinate system. To choose a coordinate system we take $ v^\mu $ as one coordinate vector corresponding to a timelike coordinate $ \tau $, and define a second coordinate vector proportional to $ u^\mu $. The newly defined coordinate vector $ \epsilon u^\mu $ is taken to correspond to a spacelike coordinate $ \sigma $. Since the time components $ v^0 = 1 $ and $ \epsilon u^0 = 0 $, these coordinates are adapted to the time slices in the target space.

The condition that these be coordinate vectors is expressed as,
\begin{align}
v^\mu \partial_\mu \epsilon u^\nu - \epsilon u^\mu\partial_\mu v^\nu =0. \label{coordinate}
\end{align}
Defining $\dot{\epsilon}$  as the $ \tau $ derivative $ v^\mu \partial_\mu \epsilon$, we can express this in terms of $ A^\mu $ and $ B^\mu $, 
\begin{align}
A^\mu \partial_\mu B^\nu - B^\mu\partial_\mu A^\nu   =  - \left ( B^\nu - A^\nu \right) \frac{\dot{\epsilon}}{\epsilon}. \label{subtraction2}
\end{align}

We can transform the equations of motion to the time slice coordinate system by using the fact that both sets of vector fields along $ A $ and $ B $ are proportional,
\begin{align}
\F{A}^\mu = \F{A}^0 A^\mu\nonumber\\
\F{B}^\mu = \F{B}^0 B^\mu.
\end{align}
Using this to expand the equations of motion \eqref{eqconformal}:
\begin{align}
A^\mu\partial_\mu B^\nu + A^\mu \Gamma^\nu_{\mu\lambda} B^\lambda + A^\mu \frac{\partial_\mu \F{B}_0}{\F{B}_0} B^\nu =0 \nonumber\\
B^\mu\partial_\mu A^\nu + B^\mu \Gamma^\nu_{\mu\lambda} A^\lambda + B^\mu \frac{\partial_\mu \F{A}_0}{\F{A}_0} A^\nu =0. \label{timeslice}
\end{align}
Note that $ A^0 = B^0 = 1 $ by definition, and thus the $\nu=0$ components of these equations imply,
\begin{align}
A^\mu \frac{\partial_\mu \F{B}_0}{\F{B}_0} = B^\mu \frac{\partial_\mu \F{A}_0}{\F{A}_0} =  -\Gamma^0_{\lambda\mu} B^\lambda A^\mu \label{relation}.
\end{align}

And so the equations of motion \eqref{timeslice} become,
\begin{align}
B^\lambda\partial_\lambda A^\nu &= -\Gamma^\nu_{\lambda\mu} B^\lambda A^\mu +\Gamma^0_{\lambda\mu} B^\lambda A^\mu A^\nu\nonumber\\
A^\lambda\partial_\lambda B^\nu &= -\Gamma^\nu_{\lambda\mu} A^\lambda B^\mu +\Gamma^0_{\lambda\mu} A^\lambda B^\mu B^\nu.\label{eqtimeslice}
\end{align}
Subtracting the two equations and comparing with the differential equation \eqref{subtraction2} for $ \epsilon $,

\begin{align}
\frac{\dot{\epsilon}}{\epsilon} = -\Gamma^0_{\lambda\mu} B^\lambda A^\mu.
\label{epdot}
\end{align}

\subsection{Worldsheet Measures}

The differential equation \eqref{epdot} does not define $ \epsilon $ uniquely, so we again have the problem of gauge dependence. Consider the gauge invariant energy-momentum tensor \eqref{T} in time slice coordinates:
\begin{align}
\tilde{T}^{\mu\nu} &= (v^\mu v^\nu - u^\mu u^\nu)\,\epsilon \,d\tau d\sigma\\
&= A^{(\mu}B^{\nu)}\,\epsilon\,d\tau d\sigma.
\label{Tepsilon}
\end{align}
Here the factor of $ \epsilon\,d\tau d\sigma $ is itself gauge invariant, being the 00-component of the energy-momentum tensor. So in any coordinate system on the worldsheet,
\begin{align}
\tilde{T}^{\mu\nu} = A^{(\mu}B^{\nu)}\,\tilde{T}^{00}.
\end{align}
 So the full energy-momentum tensor at any point can be reconstructed from the energy density $ \tilde{T}^{00} $ and the rays $ A $ and $ B $.
 
 The energy density may be gauge invariant in the sense of not depending on the worldsheet coordinates, but it depends on the coordinates in the target space. We can construct a covariant volume form $ \tilde{\omega} $ by contracting the energy-momentum tensor,
 \begin{align}
 \tilde{\omega}\equiv \frac{1}{2}\tilde{T}^{\lambda}_{\phantom{\lambda}\lambda}  
 \end{align}
 
  This turns out to just be the ordinary volume-form induced by $ h_{ab} $ which appears in the Nambu-Goto action. Using light-cone coordinates, the determinant $ h = -(\F{A}^\lambda \F{B}_\lambda)^2 $. And so using \eqref{Tconformal},
 \begin{align}
\tilde{\omega}	= \F{A}^\lambda \F{B}_\lambda \,d^2\zeta = \sqrt{-h}\,d^2\zeta
 \end{align}
 So the energy density along with $ A $ and $ B $ is enough information to construct the area of the worldsheet --- which may at first seem an entirely different measure.
 
 One more way of expressing the worldsheet area will be useful in considering intercommutations in the string network,
 \begin{align}
 |A\wedge B|\,\epsilon\,d\tau d\sigma &= \sqrt{(A^\mu B^\lambda-B^\mu A^\lambda)(A_\lambda B_\mu-B_\lambda A_\mu)}\,\epsilon\,d\tau d\sigma\nonumber\\
  &= 2\sqrt{2}\,\tilde{\omega}\label{wedge}
 \end{align}
 
\section{Kinetic Theory}\label{Sec:Kinetic}
  
Given a network of many interacting strings, the energy density can be used to form coarse-grained fields. About each point $ x $ we choose a spacetime volume $ \Delta V $ and integrate the energy density over all enclosed worldsheet area. The total energy within the volume is denoted $ \rho(x) \Delta V $. Here the density $ \rho $ is taken to be independent of the particular choice of $ \Delta V $, as long as it is chosen from an appropriate coarse-graining scale.

Other coarse-grained fields can be formed by integrating functions of $ A $ and $ B $ over the enclosed worldsheet area using the energy density as a measure. Given a function $ g $, we denote the coarse-grained field as $ \langle g \rangle $. In particular, the energy-momentum tensor of the coarse-grained string network is found by integrating \eqref{Tepsilon},

\begin{align}
T^{\mu\nu} = \langle A^{(\mu} B^{\nu)} \rangle.
\end{align}
Note that in this notation $ \langle A^0 B^0\rangle = \langle 1\rangle = \rho $.

We can also define the antisymmetric tensor $ F^{\mu\nu} $,
 \begin{align}
 F^{\mu\nu} \equiv \langle A^{[\mu} B^{\nu]} \rangle
 \end{align}
 Just as the energy-momentum tensor is conserved, it can be shown \cite{Fluid} that the $ F^{\mu\nu} $ tensor is conserved,
 \begin{align}
 \nabla_\nu T^{\mu\nu} = 0\label{consT},\\
 \nabla_\nu F^{\mu\nu} = 0\label{consF}.
 \end{align}
 The conservation of the $ F^{\mu\nu} $ tensor is related to the continuity of the strings in the network. In particular, in flat spacetime the $ \mu = 0 $ component of \eqref{consF} gives,
 \begin{align}
 \nabla \cdot \langle \B{u} \rangle = 0,\label{consU}
 \end{align}
 which is related to the fact that any string entering a volume must leave at some point.
 
 The conservation equations \eqref{consT} and \eqref{consF} can also be written in terms of $ A $ and $ B $,
 \begin{align}
\nabla_\nu \langle A^\mu B^\nu \rangle = \nabla_\nu \langle B^\mu A^\nu \rangle = 0\label{cons}.
 \end{align}

 \subsection{Distribution Function}

The conservation equations (\ref{consT}) and (\ref{consF}) constrain the dynamics of string fluid but do not describe their evolution towards equilibrium. To study the equilibration it is convenient to describe the strings in the context of the kinetic theory \cite{OldKinetic, Kinetic}. The main idea is to first derive the evolution equation (known as a transport equation) for the energy distributed over the space of all possible light-rays $ A $ and $ B $. Then, rather than integrating over the string network for each new function, we can calculate the coarse-grained fields (e.g. $\langle A^{\mu} B^{\nu} \rangle$) directly from the distribution of energy.

At any point $ x $ of the spacetime manifold $ \F{M} $, the space of all possible combinations of $ A $ and $ B $ is homeomorphic to $ S^2 \times S^2 $. We can compare $ A $ and $ B $ at different spacetime points by choosing a family of mappings from $ \Omega\equiv S^2\times S^2 $ to the product space of null rays at each $ x $. In the conformally flat spacetimes dealt with here it will not be necessary to consider these mappings explicitly. But it will still be useful to consider $ \Omega $ as a space independent of any particular spacetime point.

Given a volume $ \Delta V $ about $ x $, we can restrict the integration of the enclosed energy density to regions of the worldsheets on which $ A $ and $ B $ fall within a small interval $ \Delta\Omega $ about $ (A^\prime,B^\prime) $. The enclosed energy is then defined as $ f(A^\prime,B^\prime,x)\Delta\Omega\Delta V $. As before, in the coarse-graining approximation we take $ f(A^\prime,B^\prime,x) $ to be independent of $ \Delta \Omega $ and $ \Delta V $, which can be considered infinitessimal. 

Since $ f $ is a distribution over $ \Omega\times\F{M} $, its numerical value will depend on the measure of integration on $ \Omega $. We can map the submanifold $ S^2 $ corresponding to $ A $ (the \emph{$A$-sphere}) to the three spatial coordinates $ (A^1,A^2,A^3)\in\mathbb{R}^3$, and likewise for $ B $. This embedding of $ \Omega $ in $ \mathbb{R}^6 $ defines a measure through the pullback operation. This embedding will be useful when considering gravitational effects, but most of the results in this paper do not depend on the choice of measure.

Given $ f $, a quantity $ Q(A,B) $ coarse-grained over the string network can be calculated as an integral,
\begin{align}
\langle Q \rangle(x) = \int Q(A,B) f(A,B,x)\, d\Omega.
\end{align}
So all of the conservation equations can be thought of in terms of moments of $ f $. In particular, the requirement that the divergence of $ \langle \B{u} \rangle $ vanishes \eqref{consU} imposes a constraint on possible initial conditions for $ f $. This constraint will later be crucial to constructing the transport equation.

The energy distribution function $ f $ can also be taken as a probability distribution. Consider dividing the worldsheet area within a coarse-grained volume into small patches of equal energy. The probability that the $ (A,B) $ rays of a randomly chosen patch are in a set $ X\subseteq\Omega $ is just
\begin{align}
\frac{1}{\rho}\int_X f(A^\prime,B^\prime)d\Omega^\prime.\label{f_prob}
\end{align}
If the strings in the network are treated as random walks described by this probability distribution (in a sense to be clarified below), we can construct a transport equation for the evolution of $ f $.

\begin{figure}[]
\includegraphics[width=0.75\textwidth]{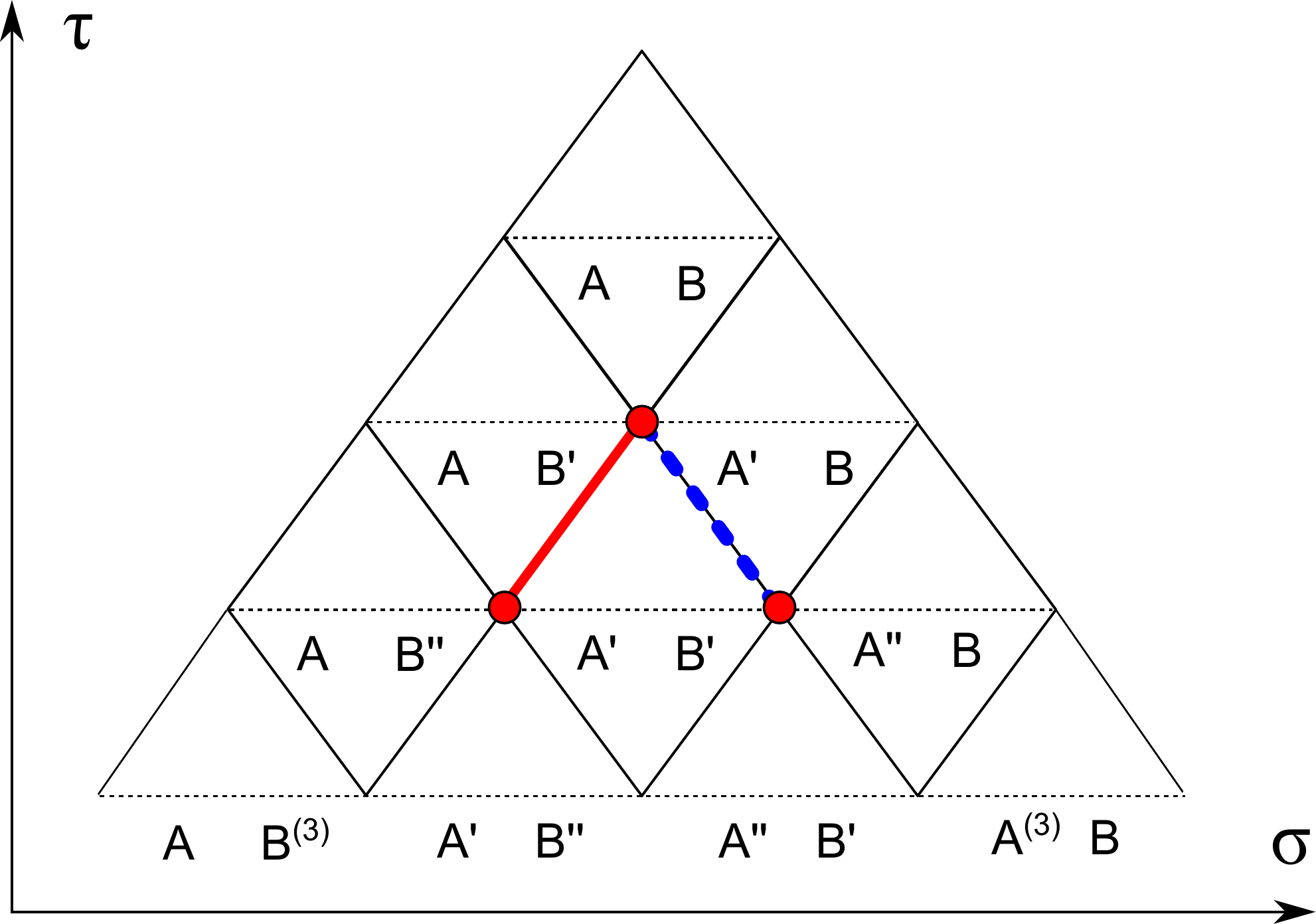}
\caption{The worldsheet of a piecewise linear string. The horizontal dotted lines represent time slices at which each segment has coordinate length $ \Delta \sigma $. The values of $ A $ and $ B $ are constant within each diamond-shaped worldsheet patch. The values of $ A $ move to the right along the string (as represented by the thick line), and the values of $ B $ move to the left (as represented by the thick dotted line). As we trace the history of a segment $ (A,B) $, we encounter other arbitrary values of $ A $ and $ B $ denoted by primes.\label{diamond}}
\end{figure}

The idea of dividing a worldsheet into patches can be made precise for strings which are piecewise linear paths. In the case of flat spacetime, a string composed of linear segments of equal energy $ \Delta\sigma $ at some initial time will continue to be piecewise linear for all time. Moreover, at time step intervals of $ \Delta\tau = \Delta\sigma/2 $, the energy of all segments will return to $ \Delta\sigma $. As shown in Fig. \ref{diamond}, each segment falls within a diamond-shaped worldsheet patch on which $ A $ and $ B $ are constant.

\subsection{Homogeneous Transport Equation}

To construct a transport equation, consider each segment in the string network as an independent entity that interacts with other segments within the coarse-graining volume. If two segments are next to each other on the same string they will interact through Nambu-Goto dynamics, and even if they are not on the same string they may interact through intercommutation. To emphasize the similarities between these interactions, they will be called \emph{longitudinal} and \emph{transverse} collisions, respectively.

To determine how a given segment $ (A,B) $ interacts through transverse collisions we must have some information on the probability distribution of segments it interacts with in the coarse-graining volume. Much like the \emph{molecular chaos} assumption made in the kinetic theory of particles, we make the assumption that the probability distribution for nearby segments is statistically independent of $ (A,B) $, and so just given by \eqref{f_prob}. We will further make a similar assumption with respect to longitudinal collisions: the state $ (A,B) $ of segment is statistically independent of the adjacent segments on the same string. In a similar model in an earlier paper \cite{OldKinetic}, these assumptions were referred to as the \emph{string chaos} assumption. We will use this term here specifically to contrast with the more general case where adjacent segments may be correlated.

We will first consider the case in flat spacetime where $ f(A,B,t) $ has no spatial dependence. As depicted in Fig. \ref{diamond}, a segment $ (A,B) $ is formed from adjacent segments $ (A,B^\prime) $ and $ (A^\prime,B) $ in the previous timestep, where $ A^\prime $ and $ B^\prime $ are arbitrary. So the total energy of segments $ (A,B) $ is equal to the total energy of segments $ (A,B^\prime) $ times the probability that the adjacent segment is $ (A^\prime,B) $. Using the string chaos assumption,
\begin{align}
f(A,B,t+\Delta t) = \int d\Omega^\prime\, f(A,B^\prime,t)\frac{f(A^\prime,B,t)}{\rho(t)}.\label{difference}
\end{align}
Counting the energy as instead coming from the segment $ (A^\prime,B) $ leads to the same equation.

Expanding to linear order in time,
\begin{align}
 \left (\der{f}{t} \right )_\text{collision} = \frac{1}{\rho}\int d\Omega^\prime\, \Gamma\cdot[ f(A,B^\prime)f(A^\prime,B)-f(A,B)f(A^\prime,B^\prime)]\label{transport_1}.
\end{align}
The subscript indicates that these are the basic collision terms in the transport equation --- later we will consider additional effects. Here the factor $ \Gamma = 1/\Delta t $ expresses the rate of the longitudinal collisions, and it is related to the correlation length (here the worldsheet coordinate length $ \Delta\sigma $) along the string. This collision rate factor $ \Gamma $ will be modified when transverse collisions are taken into account.

Under transverse collisions a string intercommutes with another string, changing the adjacent segments next to the intersection point. In principle, these intercommutations will disrupt the property that the string is linear over a segment length of $ \Delta \sigma $. But $ \Delta\sigma $ is taken to be on the order of the correlation length at equilibrium, so we continue to use the piecewise linear approximation underlying the derivation of \eqref{transport_1}.

Transverse collisions will contribute to the energy density $ f(A,B) $ on the next time step when a new sequence of segments $ (A,B^\prime) $ and $ (A^\prime,B) $ is formed, and will decrease when an existing $ (A,B) $ is disrupted by an arbitrary $ (A^\prime,B^\prime) $. These are just the two terms on the right-hand side of \eqref{transport_1}, so we just expect transverse collisions to add a contribution $ \Gamma_\perp $ to the collision rate $ \Gamma $ within the integrand. Clearly the collision rate depends on the energy density of both colliding segments, so $ \Gamma_\perp \propto \rho $. But $ \Gamma_\perp $ also depends on the orientation and relative velocity of the segments. In fact it should be proportional to the magnitude of worldsheet area \eqref{wedge} of both interacting diamonds $ |A\wedge B| $ and $ |A^\prime\wedge B^\prime| $. But if $ A\wedge B $ and $ A^\prime\wedge B^\prime $ are linearly dependent ---for instance when the string segments are pointing in the same direction--- then it should vanish. So then,
\begin{align}
\Gamma_\perp &\propto |A\wedge B \wedge A^\prime\wedge B^\prime|\\
		&\propto |(v^\prime - v)\cdot(u^\prime\times u)|,
\end{align}
where the second line is rewritten in terms of the three-velocities and tangent vectors \eqref{vu} of the interacting segments.

So denoting the proportionality constant as $ p $ (which may also include the inter-commutation probability if it is not one), the transverse collisions are taken to modify the collision rate of the transport equation \eqref{transport_1},
\begin{align}
\Gamma = \frac{1}{\Delta t} + p \rho|A\wedge B^\prime \wedge A^\prime\wedge B|. \label{Gamma}
\end{align}
Note that (\ref{transport_1}) together with (\ref{Gamma}) is in agreement with the homogeneous transport equation derived in Ref. \cite{Kinetic}.

\section{Equilibrium Distribution}\label{Sec:Equilibrium}

In a previous paper \cite{Kinetic}, it was shown that the transport equation \eqref{transport_1} implies that $ f $ converges to an equilibrium state in which the statistics of $ A $ and $ B $ are independent.
\begin{align}
\lim\limits_{t\rightarrow\infty}f(A,B,t) =  \frac{1}{\rho}\int d\Omega^\prime\, f(A,B^\prime)f(A^\prime,B)\label{chaosLim}
\end{align}
The right-hand side is a constant of the motion, and so by choosing initial conditions it can be set equal to an arbitrary factorizable $ f_\text{eq}(A,B) = f_A(A)f_B(B) $. So this form of transport equation need not lead to convergence to something analogous to the Maxwell-Boltzmann distribution --- in particular this need not converge to the Von Mises-Fisher distribution discussed in \cite{Kinetic}, although it might be a useful approximation.

However, the independence of $ A $ and $ B $ under equilibrium is a useful result which can be used to simplify the fluid equations constructed from \eqref{cons} (see Ref. \cite{Fluid} for details). The factorizable fluid model can be treated as a generalization of models of a string dust \cite{Dust}, and can also be used to describe strings with small scale structure \cite{Wiggles}, as will be discussed in a future paper \cite{NewFluid}.

The question may arise whether the independence of $ A $ and $ B $ under equilibrium depends on the assumption of string chaos. It is not an innocent assumption to take each segment of length $ \Delta\sigma $ to be completely uncorrelated to its neighbors. To relax this assumption, we might take the energy distribution of each segment to depend only on its nearest neighbors. (We will refer to this property as Markov.) 
The energy distribution over $ (A,B) $ given that the neighbor on the right is $ (A^\prime,B^\prime) $ is denoted $ f(A B \,| A^\prime B^\prime) $. Unlike the distribution $ f(A,B) $, which can be defined in general, the conditional distribution implicitly depends on the choice of $ \Delta\sigma $. Relating the two distributions,
\begin{align}
f(A,B) = \int d\Omega^\prime f(A B \,| A^\prime B^\prime) =\int d\Omega^\prime f(A^\prime B^\prime\,|A B).\label{markov}
\end{align}

Considering Fig. \ref{diamond}, the energy associated to the sequence of two segments $ (A B^\prime\,|A^\prime B) $ came from the sequence of three segments $ (A B^{\prime\prime}\,|A^\prime B^\prime\,|A^{\prime\prime} B) $ on the previous timestep, where the double primed variables are arbitrary. Due to the Markov property, the energy associated to $ (A B^{\prime\prime}) $ given $ (A^\prime B^\prime\,|A^{\prime\prime} B) $ is just $ f(A B^{\prime\prime}\,|A^\prime B^\prime) $ times the probability of the sequence $ (A^\prime B^\prime\,|A^{\prime\prime} B) $. This can written as the longitudinal part of a transport equation,
\begin{align}
f(A B^{\prime}\,|A^\prime B, t+\Delta t)= \int d\Omega^{\prime\prime} f(A B^{\prime\prime}\,|A^\prime B^\prime)\frac{f(A^\prime B^\prime\,|A^{\prime\prime} B)}{f(A^\prime, B^\prime)}.\label{transportMarkov}
\end{align}

Extending this notation, the transport equation over $ n $ steps can be written,
\begin{align}
f(A B^{\prime}\,|A^\prime B, t+n\Delta t)= \int d\Omega^{\prime\prime}\dots d\Omega^{(n+1)} f(A B^{(n+1)}\,|A^\prime B^{(n)}\,|\dots|A^{(n)} B^\prime\,|A^{(n+1)} B) \nonumber\\
= \int d\Omega^{\prime\prime}\dots d\Omega^{(n+1)} f(A B^{(n+1)}\,|A^\prime B^{(n)})\frac{f(A^\prime B^{(n)}\,|A^{\prime\prime}B^{(n-1)})}{f(A^\prime, B^{(n)})}\dots \frac{f(A^{(n)} B^{\prime}\,|A^{(n+1)} B)}{f(A^{(n)}, B^{\prime})}.
\label{transportN}
\end{align}

The repeated product has the form of a right stochastic matrix,
\begin{align}
F^{AB}_{\quad A^\prime B^\prime} \equiv \frac{f(A B\,|A^{\prime} B^\prime)}{f(A, B)}.
\end{align}
This matrix has a normalized right eigenvector equal to $ f(A, B)/\rho $, as can be verified using \eqref{markov}. So upon repeated matrix multiplication, this matrix converges to a matrix $ G $ with each row equal to $ f(A, B)/\rho $. Thus we can find the limit of \eqref{transportN},
\begin{align}
\lim\limits_{t\rightarrow\infty} f(A B^{\prime}\,|A^\prime B, t)&=\int d\Omega^{(2)}d\Omega^{(3)} f(A B^{(3)}\,|A^\prime B^{(2)})\,G^{A^\prime B^{(2)}}_{\quad A^{(2)} B^{\prime}}\,\frac{f(A^{(2)} B^{\prime}\,|A^{(3)} B)}{f(A^{(2)}, B^{\prime})}\nonumber\\
&= \frac{1}{\rho}\int d\Omega^{(2)}d\Omega^{(3)} f(A B^{(3)}\,|A^\prime B^{(2)})f(A^{(2)} B^{\prime}\,|A^{(3)} B)
\label{markovLim}
\end{align}

So the limit does not reduce to the string chaos assumption. But the distribution once again factors into independent distributions over $ A $ and $ B $. In fact, if we integrate over $ \Omega^\prime $ we see that $ f(A,B) $ converges to the same limit as the string chaos case \eqref{chaosLim}. This transport equation \eqref{transportMarkov} did not take the transverse collisions into account. But since the transverse collisions do not depend on correlations of nearby segments, we would not expect the inclusion of the transverse collisions to lead to a non-factorizable equilibrium.

Finally, note that even if the probability distribution depends on the $ n $ nearest segments, we would still expect the factorizability to hold. The longitudinal term in this case would be,
\begin{align}
f(A B^{(n)}\,|A^\prime B^{(n-1)}\,&|\dots|A^{(n-1)} B^\prime\,|A^{(n)} B, t+\Delta t) \nonumber
\\&= \int d\Omega^{(n+1)}\,f(A B^{(n+1)}\,|\dots|A^{(n)} B^\prime)
\frac{f(A^\prime B^{(n)}\,|\dots|A^{(n+1)} B)}{\hat{f}(A^\prime B^{(n)}\,|\dots|A^{(n)} B^\prime)},\label{transportUgly}
\end{align}
where $ \hat{f} $ is a energy distribution over $ n-1 $ segments, satisfying a property like \eqref{markov}.

Suppose that $ f $ can be factored as $ f = f_A\,f_B $, and thus $ \hat{f} = \hat{f}_A\,\hat{f}_B $. The $ f_B $ from the first factor in the integrand contains an argument $ B^{(n+1)} $ which is integrated over. Likewise for the $ f_A $ from the second factor. Together, these cancel with the $ \hat{f} $ in the denominator, and the remaining $ f_A $ and $ f_B $ have arguments in the same order as on the left-hand side. Thus any factorizable $ f $ is a fixed point of the transport equation \eqref{transportUgly}, and we would conjecture conversely that an equilibrium has the factorizability property.

\section{Inhomogeneous Transport Equation}\label{Sec:Inhomogeneous}

The spatially homogenous transport equation satisfies the conservation equations \eqref{cons} with all spatial derivatives set to zero --- i.e. $ \rho $ and the components $ \langle A^i\rangle $ and $ \langle B^i\rangle $ are all time independent. The most straightforward way to extend the transport equation to $ f(A,B,x) $ is to take the energy associated with segments $ (A,B) $ to move through space with velocity $ v = (A+B)/2 $. This agrees with the approach taken for particles and this was the approach taken in \cite{Kinetic}. The resulting transport equation does lead to the proper conservation equation for energy,
\begin{align}
\der{\rho}{t} = -\nabla \cdot \langle \B{v}\rangle,
\end{align}
But the conservation equations describing the spatial transport of $ \langle A^i\rangle $ and $ \langle B^i\rangle $ are not respected.

The problem is that the equations of motion \eqref{eqconformal} imply the quantities $ A^i $ move through space in the direction of $ B $, whereas the quantities $ B^i $ move in the direction of $ A $. To enforce this situation, we can consider the segment $ (A,B) $ to describe a particle carrying energy and charge $ A $ which moves with velocity $ B $ in between collisions. These \emph{$A$-particles} emit virtual \emph{$B$-particles} that move with velocity $ A $. In a collision, a virtual particle is absorbed and the $ B $ velocity is changed.

So the energy associated to $ (A,B) $ at a given location $ x $ came from the energy of $A$-particles $ (A,B^\prime) $ at locations $ x - B^\prime\Delta t $. The only $A$-particles that contribute are those that collide with $B$-particles emitted by some $ (A^\prime,B) $ at a location $ x - A^\prime\Delta t $. This picture of independently moving $A$ and $B$ charges can be thought of instead in terms of the lower vertex of the worldsheet diamonds in Fig. \ref{diamond}. The $A$-particles are represented as circles, and their path by the thick line. The path of the virtual $B$-particle is represented by the thick dotted line.

Using this picture we can form a difference equation much like in the homogenous case:
\begin{align}
f(A,B,x,t+\Delta t) = \int d\Omega^\prime\, f(A,B^\prime,x-B^\prime\Delta t,t)\frac{f(A^\prime,B,x-A^\prime\Delta t,t)}{\rho_B}.\label{transport_2diff}
\end{align}
Here $ \rho_B $ is the total energy density at $ x $ associated with $B$-particles, and is used to normalize the probability of colliding with a particle emitted by $ (A^\prime,B) $. Calculating $ \rho_B $ to linear order in $ \Delta t $,
\begin{align}
\rho_B &= \int d\Omega^{\prime\prime} f(A^{\prime\prime},B^{\prime\prime},x-A^{\prime\prime}\Delta t,t)\nonumber\\
		&= \rho - \nabla\cdot\langle \B{A}\rangle\,\Delta t
\end{align}
But due to the constraint derived from the continuity of strings \eqref{consU},
\begin{align}
\nabla\cdot\langle \B{A}\rangle = \nabla\cdot\langle \B{B}\rangle = \nabla\cdot\langle \B{v}\rangle.\label{consU2}
\end{align}
So we could instead consider the energy to flow with the $B$-particles and obtain the same difference equation \eqref{transport_2diff}.

The transverse collisions can be taken to happen at the same point in space and so need not introduce any additional terms into the spatially dependent transport equation. So expanding the difference equation to first order in $ \Delta t $, we again find the collision terms \eqref{transport_1}, but now there are additional spatial terms,
\begin{align}
\left (\der{f}{t} \right )_\text{spatial} = - \frac{1}{\rho}\int d\Omega^\prime\, f(A,B^\prime)\overleftrightarrow{\nabla}f(A^\prime,B),\label{transport_2}
\end{align}
where the operator $ \overleftrightarrow{\nabla} $ is defined to act on both the left and right,
\begin{align}
\overleftrightarrow{\nabla} \equiv \overleftarrow{\nabla}\cdot\B{B}^\prime + \B{A}^\prime\cdot\overrightarrow{\nabla} - \frac{1}{\rho}\nabla\cdot\langle \B{v}\rangle \label{arrowsnabla}.
\end{align}

To check that this respects the conservation equations we can multiply both sides of the transport equation by 1, $ A^i $, or $ B^i $ and integrate over $ \Omega $.  The collision term on the right-hand side integrates to zero \cite{Kinetic}, and using the constraint \eqref{consU2} we find the correct conservation equations,
\begin{align}
\der{}{t}\rho + \partial_k\langle v^k \rangle = 0 \label{rhoeq}\\ 
\der{}{t}\langle A^i \rangle + \partial_k\langle A^i B^k\rangle = 0 \label{Aeq}\\
\der{}{t}\langle B^i \rangle + \partial_k\langle B^i A^k\rangle = 0 \label{Beq}
\end{align} 
in agreement with (\ref{cons}).

However one remaining difficulty of these transport equations \eqref{transport_2} is that they are not covariant. Under a coordinate transformation, spatial derivatives appear on the left-hand side and time derivatives appear within the operator $ \overleftrightarrow{\nabla} $. A possible alternate approach is to instead take the time difference to appear on the right-hand side of \eqref{transport_2diff} as $ t-\Delta t $. The resulting transport equation would also respect the conservation equations and would be Lorentz invariant (more general transformations are discussed below). However, the time derivatives only appear within the integral terms and the transport equation would not uniquely specify $ f $ given an initial condition.

\section{Friedmann Universe}\label{Sec:Friedmann}

So far the transport equation has been constructed assuming the energy and $ A $ and $ B $ vectors are conserved in the time interval between collisions. In the presence of an external field this is no longer true. Given a nontrivial metric, the connection coefficients in the equations of motion \eqref{eqtimeslice} and \eqref{epdot} introduce gravitational corrections. The corrections depend on $ A $ and $ B $, but not any higher-order derivatives on the worldsheet. So we can account for the effect of the gravitational field through additional terms in the transport equation for $ f(A,B,x) $.

A simple and physically relevant case is that of the conformally flat Friedmann metric,
\begin{align}
ds^2 = a^2(\tau)(\tau^2 - \B{x}^2)\label{metric}
\end{align}
In this case, the nonzero connection coefficients are all equal to $ \Hub \equiv \dot{a}/a $. The change in energy density \eqref{epdot} reduces to,
\begin{align}
\frac{\dot{\epsilon}}{\epsilon} = -\Hub(1+\B{A}\cdot\B{B})\label{epfriedmann}.
\end{align}
And the equations of motion \eqref{eqtimeslice} become,
\begin{align}
B^\mu\partial_\mu{A^i}&=-\Hub(B^i-(\B{A}\cdot\B{B})A^i)\nonumber\\
A^\mu\partial_\mu{B^i}&=-\Hub(A^i-(\B{A}\cdot\B{B})B^i).
\label{eqfriedmann}
\end{align}

Since $ f(A,B) $ is proportional to $ \epsilon $, \eqref{epfriedmann} will introduce a term into the transport equation representing the overall change in energy. Accounting for this and the change in $ A $ and $ B $, we can modify the left-hand side of the difference equation $ \eqref{difference} $,
\begin{align}
(1-\frac{\dot{\epsilon}}{\epsilon}\Delta t)\,f(A+\Delta A,B+\Delta B,t+\Delta t)\,d\Omega^\prime\,dV^\prime.
\label{diffFriedmann}
\end{align}
Here $ \Delta A $ and $ \Delta B $ are the changes in the coordinates on $ \Omega $ due to \eqref{eqfriedmann}. Now we must also consider the changes in the volume elements with time. Expanding to linear order in $ \Delta t $, we will find five new terms in the transport equation --- all proportional to $ \Hub $. The three terms resulting from the changes in $ \epsilon $, $ dV $, and $ d\Omega $ will all be proportional to $ f $. The two terms due to $ \Delta A $ and $ \Delta B $ will result in derivatives of $ f $ with respect to the coordinates on $ \Omega $. These terms (with $ \Hub $ factored out) will be written as $ \partial_A f$ and $ \partial_B f$.

As discussed in connection with the time derivative, all of these corrections could instead be implemented on the right-hand side of the difference equation (\ref{difference}). This would lead to new integro-differential terms appearing in the operator $ \overleftrightarrow{\nabla} $. We will instead continue to consider the model described by \eqref{diffFriedmann}.

\subsection{Differential Terms}\label{differential}
In the following discussion we will focus on the $A$-sphere in $ \Omega $. Everything will extend to the $B$-sphere in the obvious way. Using \eqref{eqfriedmann}, the derivative $ \partial_A $ can be written formally in terms of the embedding of the $A$-sphere in $ \mathbb{R}^3 $:
\begin{align}
\partial_A\equiv (\B{B}-(\B{A}\cdot\B{B})\B{A})\cdot\der{}{\B{A}}.\label{dA}
\end{align}
Of course the three components $ A^i $ do not form a coordinate system on the two-dimensional $A$-sphere. But note that the vector $ \B{B}-(\B{A}\cdot\B{B})\B{A} $ is always orthogonal to the unit vector $ \B{A} $, and so the vector is in the tangent space of the embedded $A$-sphere. Thus this derivative in $ \mathbb{R}^3 $ can be considered a push-forward of a proper two-dimensional derivative. This expression itself will be useful in taking derivatives of functions of $ A^i $.

For a proper two-dimensional coordinate system, we can take $ \B{B} $ as the z-axis of a polar coordinate system $ (\theta_A,\phi_A) $ on the $A$-sphere. In terms of these coordinates,
\begin{align}
\partial_A = -\text{sin}\,\theta_A\der{}{\theta_A}.\label{dTheta}
\end{align}
For completeness, note that the $ \phi_A $ derivative also has a simple form,
\begin{align}
\der{A^i}{\phi_A} = (\B{B}\times\B{A})^i.
\end{align}

In a practical simulation of the transport equation it may be useful to use coordinate systems that do not vary with $ \B{B} $. In terms of the three fixed polar coordinate systems $ (\theta_A^i,\phi_A^i) $ about the coordinate axes in $ \mathbb{R}^3 $, it is easy to show,
\begin{align}
\partial_A = -\sum_i \,B^i\, \text{sin}\,\theta_A^i\,\der{}{\theta^i_A}
\end{align}

\subsection{Measure Terms}
In terms of a general coordinate system $ \xi^a $ on $ \Omega $, the measure $ d\Omega = \omega(\xi)\,d\xi^1\dots d\xi^4$, for some distribution $ \omega $. Due to the flow \eqref{eqfriedmann}, the coordinates $ \xi^a $ change to $ \xi^{a^\prime} = \xi^a + \Delta\xi^a $. The transformed volume element $ d\Omega^\prime $ can be written,
\begin{align}
d\Omega^\prime &= \omega(\xi+ \Delta\xi^a)\det\left\lvert\der{(\xi^{1^\prime}\dots \xi^{4^\prime})}{(\xi^1\dots \xi^4)}\right\rvert\, d\xi^1\dots d\xi^4.
\end{align}
To first order, only the diagonal terms in the Jacobian contribute:
\begin{align}
d\Omega^\prime &= (\omega+\Delta\xi^a \partial_a \omega) (1 + \der{\Delta\xi^a}{\xi^a})\, d\xi^1\dots d\xi^4\nonumber\\
				&= (1 + \frac{1}{\omega}\Delta\xi^a \partial_a \omega+ \ \der{\Delta\xi^a}{\xi^a} )\,d\Omega.
\end{align}

These correction terms can be calculated rigorously using the polar coordinate systems above. For a simple derivation, we will treat the two Euclidean coordinates in $ \mathbb{R}^3 $ which are perpendicular to $ A^i $ as coordinates on the $A$-sphere. So $ d\Omega = dA^1\, dA^2\, dB^1\, dB^2 $, and using \eqref{eqfriedmann} the correction terms become,
\begin{align}
d\Omega^\prime &= (1 + \sum_{i=1,2}\left. \der{\Delta A^i}{A^i}\right\rvert_{A^i=0}+ \sum_{i=1,2} \left.\der{\Delta B^i}{B^i}\right\rvert_{B^i=0})\,d\Omega\nonumber\\
&= (1 - \sum_{i=1,2}\left. \der{}{A^i}\Hub(B^i-(\B{A}\cdot\B{B})A^i)\Delta t\right\rvert_{A^i=0}- \dots)\,d\Omega\nonumber\\
&=(1+4\Hub(\B{A}\cdot\B{B})\Delta t)\,d\Omega.\label{dOmegaterm}
\end{align}

There is also a change in the volume element $ dV $ due to the time change in $ \sqrt{-g} $,
\begin{align}
dV^\prime &= (1+\frac{1}{\sqrt{-g}}\partial_0\sqrt{-g}\,\Delta t)dV\nonumber\\
&= (1+\Gamma^\mu_{0 \mu}\,\Delta t)dV\nonumber\\
&= (1+4\Hub\,\Delta t)dV\label{dVterm}
\end{align}
Instead we could absorb the $ \sqrt{-g} $ into the definition of $ f $, and consider $ T^{\mu\nu} $ and $ F^{\mu\nu} $ in the conservation equations \eqref{cons} as tensor densities. The conservation equations would then be modified by a term involving $ \Gamma^\mu_{\lambda \mu} $, which vanishes except for $ \lambda = 0 $. So \eqref{dVterm} is just what is needed for consistency.

\subsection{Conservation Equations}

Integrating the new transport equation  over $ \Omega $ should recover the conservation equations for a fluid of strings. The new gravitational terms should just account for the connection coefficients in the covariant derivatives in \eqref{cons}. As already noted, both the term of $ 4\Hub f $ from \eqref{dVterm} and the $ \Gamma^\mu_{\lambda \mu} $ connection coefficient can be eliminated by absorbing a factor of $ \sqrt{-g} $ into $ f $. Writing the remaining gravitational terms,
\begin{align}
(\der{f}{t})_\text{gravitational} = \Hub\,(\partial_A + \partial_B - (1 + \B{A}\cdot\B{B})-4\B{A}\cdot\B{B})\,f\label{transport_3}
\end{align}

To verify the conservation equation, we integrate the gravitational terms multiplied by a function $ Q(A,B) $:
\begin{align}
\Hub\int Q (\partial_A f+ \partial_B f)\,d\Omega - \Hub\langle Q(1 + \B{A}\cdot\B{B})\rangle- 4\Hub\langle Q(\B{A}\cdot\B{B})\rangle.\label{Qint}
\end{align}
The first term can be integrated by parts using \eqref{dTheta},
\begin{align}
\int Q\, \partial_A f\,d\Omega &= \int Q (-\text{sin}\,\theta_A\der{f}{\theta_A} )(\text{sin}\,\theta_A\,d\theta_A\,d\phi_A\cdot\text{sin}\,\theta_B\,d\theta_B\,d\phi_B)\nonumber\\
&= -\int f\, \partial_A Q\,d\Omega + \int  Qf(2\,\text{cos}\,\theta_A)d\Omega
\end{align}
And since $ \B{A}\cdot\B{B} = \text{cos}\,\theta_A = \text{cos}\,\theta_B $, the second term is $ 2\langle Q(\B{A}\cdot\B{B})\rangle $. Along with the corresponding term from the integral involving $ \partial_B $, this cancels with the final term in \eqref{Qint}. So the integral of $ Q $ times the gravitational terms becomes,
\begin{align}
-\Hub\langle \partial_A Q+ \partial_B Q\rangle- \Hub\langle Q(1 + \B{A}\cdot\B{B})\rangle\label{Qint2}
\end{align}
Choosing $ Q$ to be 1, $ A^i $, $ B^i $ and using \eqref{dA}, we see this is indeed just the gravitational correction to the conservation equations,
\begin{align}
&\partial_\nu\langle v^\nu\rangle = -\Hub\langle 1 + \B{A}\cdot\B{B}\rangle\label{eqRho}\\
&\partial_\nu\langle A^i B^\nu\rangle = -2\Hub \langle v^i\rangle\label{eqAB}\\
&\partial_\nu\langle B^i A^\nu\rangle = -2\Hub \langle v^i\rangle
\end{align}
So this transport equation \eqref{transport_3} is fully consistent with the fluid equations derived in \cite{Fluid}.

\section{Summary}\label{Sec:Summary}

The main result of the paper is a derivation of a transport equation which reduces to the transport equation in Ref. \cite{Kinetic} in the homogeneous limit and is in agreement with the fluid equations in Ref. \cite{Fluid} with spatial dependence and background gravitational effect taken into account. Schematically the transport equation can be written as
\begin{align}
\der{f}{t} = \left (\der{f}{t} \right )_\text{collision} + \left (\der{f}{t}\right )_\text{spatial} +\left (\der{f}{t}\right )_\text{gravitational} \label{transport_final}
\end{align}
where the terms are given by \eqref{transport_1}, \eqref{transport_2}, and \eqref{transport_3}, respectively. We shall now briefly review the origin of each of these terms.

The collision term defined by \eqref{transport_1} and \eqref{Gamma} represents the Nambu-Goto evolution (or longitudinal collisions)  $\propto (\rho \Delta t)^{-1}$ and inter-commutations (or transverse collisions) $\propto p |A\wedge B^\prime \wedge A^\prime\wedge B|$. The analysis of strings with Markov property in Sec. \ref{Sec:Equilibrium} had shown that the parameter $\Delta t$ should be set by the time scale of local equilibration or, equivalently, by the correlation length of strings. In principle, this should be a dynamical parameter, but since the factorizability property of an equilibrium does not depend on $\Delta t$ we do not expect the solutions to the transport equation to change significantly given that $\Delta t$ is sufficiently small. On the other hand the equilibration of the strings with both longitudinal and transverse collisions is an important open question that we are currently trying to address using numerical techniques \cite{NumericKinetic}. 

The spatial term \eqref{transport_2} is not a differential term as it is in the Boltzmann transport equation, but an integral term which was derived by expanding the integral term responsible for the longitudinal collisions. The longitudinal collisions have no analog in the case with particles and arise due to the Nambu-Goto evolution of strings as illustrated on Fig. \ref{diamond}. When the probability of a given longitudinal collision between $A^\mu$ and $B^\mu$ is calculated one should keep in mind that $A$-particles are moving with velocities $B^\mu$ and $B$-particles are moving with velocities $A^\mu$. Then the spatial term \eqref{transport_2} involves integrals over spatial gradients of $f(A,B)$, but the time derivatives remain outside of the integral. This breaks the general covariance in the inhomogeneous  transport equation, although the covariance is maintained at the level of conservation equations (\ref{rhoeq}), (\ref{Aeq}) and (\ref{Beq}) which can also be derived from a fluid perspective \cite{Fluid}.

The gravitational terms \eqref{transport_3} were calculated for a Friedmann universe, by treating the Friedmann expansion in the external field approximation (i.e. by neglecting the gravitational back-reaction). Once background gravitational effects are taken into account the transport equation gets a contribution due to the redshift of energy $-\Hub(1 + \B{A}\cdot\B{B})f$. The expansion also affects the tangent space of a worldsheet, changing $A^\mu$ and $B^\mu$. This effect leads to additional differential terms $ \partial_A f $ and $ \partial_B f $ in the transport equation, the form of which depends on the coordinates in phase-space discussed in Sec. \ref{differential}. In addition, there is a correction $-\Hub(4 \B{A}\cdot\B{B})f$ due to the non-Hamiltonian convergence of these phase-space trajectories (i.e. negative Lyapunov exponents).

\section*{Acknowledgments}

The authors are grateful to Jacob Balma for frequently being confused. This work was supported in part by the Start-Up funds (V.V.) and by the Grant-In-Aid (D.S.) from the University of Minnesota and by the Frank and Ruth Friebe Scholarship (D.S.).

\end{document}